\documentclass[a4paper,twocolumn,english,,prb,showpacs]{revtex4}
\usepackage[T1]{fontenc}
\usepackage[latin9]{inputenc}
\usepackage{amsmath}
\usepackage{amssymb}
\usepackage{graphicx}
\usepackage{esint}

\makeatletter

\providecommand{\tabularnewline}{\\}

\@ifundefined{textcolor}{}
{%
 \definecolor{BLACK}{gray}{0}
 \definecolor{WHITE}{gray}{1}
 \definecolor{RED}{rgb}{1,0,0}
 \definecolor{GREEN}{rgb}{0,1,0}
 \definecolor{BLUE}{rgb}{0,0,1}
 \definecolor{CYAN}{cmyk}{1,0,0,0}
 \definecolor{MAGENTA}{cmyk}{0,1,0,0}
 \definecolor{YELLOW}{cmyk}{0,0,1,0}
}

\usepackage{babel}

\usepackage{babel}

\makeatother

\usepackage{babel}
\begin{document}

\title{\emph{Ab initio} study of $2p$-core level x-ray photoemission spectra
in ferromagnetic transition metals}

\author{Manabu Takahashi$^{1}$ and Jun-ichi Igarashi$^{2}$ }

\affiliation{$^{1}$Faculty of Engineering, Gunma University, Kiryu, Gunma 376-8515,
Japan\\
 $^{2}$Faculty of Science, Ibaraki University, Mito, Ibaraki 310-8512,
Japan}
\begin{abstract}
We study the 2p-core level x-ray photoemission spectra in ferromagnetic
transition metals, Fe, Co, and Ni using a recently developed ab initio
method.The excited final states are set up by distributing electrons
on the one-electron states calculated under the fully screened potential
in the presence of the core hole. We evaluate the overlap between
these excited states and the ground state by using one-electron wave
functions, and obtain the spectral curves as a function of binding
energy. The calculated spectra reproduce well the observed spectra
displaying interesting dependence on the element and on the spin of
the removed core electron. The origin of the spectral shapes is elucidated
in terms of the one-electron states screening the core hole. The magnetic
splitting of the threshold energy is also estimated by using the coherent
potential approximation within the fully screened potential approximation.
It decreases more rapidly than the local spin moment with moving from
Fe to Ni. It is estimated to be almost zero for Ni despite the definite
local moment about $0.6\,\mathrm{\mu_{B}}$, in agreement with the
experiment. 
\end{abstract}

\pacs{79.60.-i 71.15.Qe 71.20.Be}

\maketitle

\section{Introduction}

Core-level x-ray photoemission spectroscopy (XPS) is one of the powerful
tools for studying the electronic structure in solids through the
response of electrons to the photocreated core hole. It is well known
that the response function in metals displays singular behavior near
the Fermi edge.\cite{Anderson1967,Mahan1967,Nozieres1969} Core-level
XPS spectra as a function of binding energy display asymmetric shape
near the threshold.\cite{Doniach1970} The spectra sometimes have
extra structures in the high binding energy region. A notable example
in metals is a satellite peak on the $2p$ XPS in Ni metal, which
appears around $6$ eV higher than the threshold.\cite{Hufner1975}
Feldkamp and Davis\cite{Feldkamp1980} analyzed the Ni $2p$ XPS spectra
by evaluating the overlap determinants between the ground and excited
states, using a numerical method on the Hubbard-like model. They clarified
the origin of satellite as a combined effect of the core-hole screening
and the interaction between electrons, and estimated the strength
of the core-hole potential consistent with the binding energy of the
satellite intensity and the asymmetry parameter.

In addition to the above features, core-level XPS spectra present
several intriguing behaviors in ferromagnetic transition metals Fe,
Co, and Ni. On the spin resolved $3s$ spectra, they show characteristic
satellite intensities in the majority spin channel for all Fe, Co,
and Ni, while they show almost single peak structures in the minority
spin channel. In Fe, the spectra exhibit particularly a large satellite
structure only in the majority spin channel. In contrast to the $3s$
spectra, the $2p$ spectra do not have a clear satellite peak in Fe,
while they show the notable $6$-eV satellite in Ni. As regards the
peak around the threshold in $2s$, $2p$, and $3s$ spectra, its
positions noticeably depend on the spin channel in Fe, while such
dependence has not been observed in Ni.\cite{Acker1988,Campen1994}
According to the recent hard x-ray photoemission spectroscopy (HAXPES)
experiment,\cite{Imada.unpub} such magnetic splitting is estimated
about $0.9$ eV for the $2p_{3/2}$ core in ferromagnetic Fe. Similar
magnetic splittings have also been observed on the $2p_{3/2}$ spectra
in several half-metallic ferromagnetic Heusler alloys.\cite{Plogmann1999,Slebarski2001,Shukla2007,Cui2008,Gray2011}
Such splittings are usually considered to be related with the local
spin moment at the photo-excited site. As regards the satellite structure,
it has been clearly observed in the $3s$ XPS in several Fe compounds.\cite{Acker1988}
We have clarified the origin of the satellite intensity on the $3s$
spectra and that the satellite intensity is not a direct reflection
of the local spin moment by calculating the XPS spectra on the \emph{ab
initio} level. \cite{Taka2008FeXPS,Taka2010FeCoNi3sXPS}

For the strongly correlated localized electron systems such as the
$3d$ transition metal oxides and the $f$-electron systems, the theoretical
analysis of the XPS spectra have been carried out mainly on the basis
of atomic, cluster, or impurity Anderson models. \cite{deGrootKotani}
Although the impurity Anderson model has been applied to analyze the
core-level spectra in some itinerant metallic systems, \cite{Tanaka1992b,Shukla2007}
it is not suitable to analyze the spectra in the highly itinerant
metallic systems. Another approach to investigate the core-level spectra
is based on the independent one-electron theory exploiting a meanfield
approximation. Mahan,\cite{Mahan.PRB.21.1421} and Barth and Grossmann\cite{vonBarth1979}
have calculated the overall line shapes of the x-ray spectra by evaluating
one-electron wave functions under the final state potential, and have
reproduced well the experimental emission and absorption spectra in
metals within the meanfield theory.

Recently we have presented the XPS spectra based on the \emph{ab initio}
band structure calculation.\cite{Taka2008FeXPS,Taka2010FeCoNi3sXPS}
We have calculated the final state potential self-consistently by
carrying out the band structure calculation on the system holding
a core-hole at a photo-excited site. Not only the core-hole potential
but also the relaxation of the core states as well as the screening
electron distribution are automatically determined through the calculation.
Distributing electrons on the energy levels thus evaluated, we have
constructed various final states. We have evaluated the overlaps between
those final states and the ground states by using the one-electron
wave functions, and finally obtained the $3s$ core-level XPS spectra
in the ferromagnetic transition metals Fe, Co, and Ni. The element
and photo-electron-spin dependence of the spectral line shape have
been reproduced in good agreement with experiments within the\emph{
ab initio} manner.\cite{Taka2008FeXPS,Taka2010FeCoNi3sXPS} This method
may be regarded as an extension of the Feldkamp-Davis method\cite{Feldkamp1980}
to an \emph{ab initio} level. The origin of the spectral shape as
a function of the binding energy has been elucidated in terms of the
one-electron states screening the core-hole as follows.

In the fully screened state where an up-spin $3s$ electron is removed,
the down-spin $3d$ states are strongly attracted in the core-hole
site, forming quasibound states near the bottom of the $3d$ band
(hereafter majority spin and minority spin are called as up-spin and
down-spin, respectively). The final states that the quasibound states
are unoccupied contribute to the satellite or shoulder intensities.
In the state where a down-spin $3s$ electron is removed, the up-spin
$3d$ electrons are strongly attracted to the core hole, forming quasibound
states near the bottom of the $3d$ band, but the down-spin $3d$
electrons are not attracted strongly enough to form quasibound states.
Because the up-spin $3d$ states are almost fully occupied in the
ground state, the quasibound states, which appear only in the up-spin
$3d$ states, could not become unoccupied, leading to a single peak
structure of the spectra.

In this paper, we apply our method to calculate the $2p$ core-level
XPS spectra in ferromagnetic Fe, Co, and Ni, and elucidate the underlying
mechanism. The calculation could not distinguish between the $2p_{3/2}$
core and the $2p_{1/2}$ core, because the spin-orbit interaction
is not taken into account. We focus on the difference of the spectra
among the three ferromagnetic transition metals and on the difference
between the $2p$ and $3s$ spectra. The main features of the $2p$
spectra are consistently reproduced in the same manner as those of
$3s$ spectra. We could understand the origin of the spectral shape
in the high binding energy region by considering an electron-hole
pair excitation from the quasibound states with down-spin to the unoccupied
states in the fully screened state. In Ni, the fully screened one-electron
states are similar both in the $2p$ and $3s$ electron removal states,
while they are significantly different in Fe. Such different screening
behaviors of the $3d$ states cause the difference in the spectral
shapes between Fe and Ni. It is found that the band filling in the
down-spin state plays important roles to give rise to the difference
between the Fe $2p$ and Ni $2p$ spectra. We also evaluate the magnetic
splitting of the threshold energy, exploiting the coherent potential
approximation (CPA) within the fully screened potential approximation.
This may be observed as the magnetic splitting of the peak position
around the threshold in the spin-resolved XPS spectra. We obtain the
magnetic splitting of Fe $2p_{3/2}$ as $0.9$ eV in agreement with
the recent experiment, and that of Ni $2p_{3/2}$ as nearly zero despite
the finite local magnetic moment about $0.6\,\mathrm{\mu_{B}}$.

The present paper is organized as follows. In Sec. II, we briefly
describe the calculation procedure of the XPS spectra. In Sec. III,
we present the $3d$ band calculated in the presence of core hole
and the XPS spectra in comparison with the experiments. The last section
is devoted to the concluding remarks.

\section{Calculation Method}

\subsection{Calculation of spectral intensity}

We briefly summarize the calculation procedure for the core-level
photoemission spectra. For details, we refer the readers to Ref.{[}\onlinecite{Taka2010FeCoNi3sXPS}{]}.
The many-body wave functions of the ground and final states are assumed
to be given as single Slater determinants.

For the ground state, we carry out the full potential linear augmented
plane wave (FLAPW) band structure calculation based on the local density
approximation (LDA), and obtain the one-electron wave functions $\phi_{j}$'s.
We construct the Slater determinant by putting electrons from the
lowest energy state up to the Fermi level. For the final states, we
carry out the same type of band calculation under the fully screened
potential in a periodic array of supercells with one core hole per
cell. The core-states are treated as localized states within a muffin-tin
sphere, so that we could specify the core-hole site. In reality, only
one core-hole should exist in crystal through the XPS event. Therefore,
the larger the unit cell size is, the better results would be expected
to come out. We use the $3\times3\times3$ bcc supercell for Fe and
fcc supercell for Co and Ni, where the core-hole sites form a bcc
super lattice and an fcc super lattice, respectively. In the self-consistent
procedure, we keep a hole in a specified core level at a core-hole
site, and put an extra electron in each super cell to guarantee the
charge neutrality. The local charge neutrality would be satisfied,
known as the Friedel sum rule in the impurity problem.\cite{Friedel1958}
We thus obtain the one-electron wave function $\psi_{i}$ with energy
eigenvalue $\epsilon_{i}$, which takes account of not only the effect
of core-hole potential but also that of electron-electron interactions
within the limit of the LDA. The final state $\left|f\left(0\right)\right\rangle $
with the lowest energy is constructed by putting electrons from the
lowest energy state up to the Fermi level at each $\mathbf{k}$ point,
as was done in the ground state. The other final states are obtained
by creating electron-hole (e-h) pairs from this final state $\left|f\left(0\right)\right\rangle $.
We designate the state having the $\ell$ e-h pairs on the state $\left|f\left(0\right)\right\rangle $
as $\left|f\left(\ell,m\right)\right\rangle $, where the index $m$
distinguishes the different electron configuration. The overlap integral
$a_{i,j}^{\left(\ell,m\right)}$ between the wave function $\psi_{i\left(\ell,m\right)}$
for the $i$'th occupied valence state in the final state $\left|f\left(\ell,m\right)\right\rangle $
and the wave function $\phi_{j}$ for the $j$'th occupied valence
state in the ground state$\left|g\right\rangle $ is given by as $a_{i,j}^{\left(\ell,m\right)}=\int\psi_{i\left(\ell,m\right)}^{*}\phi_{j}dr^{3}$,
where the volume integral is taken over a super cell.

Neglecting the interaction between the escaping photo-electron and
the other electrons in matter, we consider the XPS process that a
core-electron is excited to a high energy state with energy $\epsilon$
by absorbing an x-ray photon with energy $\omega$. Exploiting Fermi's
golden rule, we obtain the expression of the spectral intensity as
a function of the biding energy $\omega-\epsilon$ as 
\begin{align}
I_{\sigma}^{{\rm XPS}}\left(\omega-\epsilon\right) & =A\sum_{\ell,m}\left|\begin{array}{ccc}
a_{1,1}^{\left(\ell,m\right)} & \cdots & a_{1,N_{\mathrm{e}}}^{\left(\ell,m\right)}\\
\vdots & a_{i,j}^{\left(\ell,m\right)} & \vdots\\
a_{N_{\mathrm{e}},1}^{\left(\ell,m\right)} & \cdots & a_{N_{\mathrm{e}},N_{\mathrm{e}}}^{\left(\ell,m\right)}
\end{array}\right|^{2}\nonumber \\
 & \times\delta\left(\omega-\varepsilon-E_{0}+E_{g}-\Delta E_{\left(\ell,m\right)}\right),\label{eq:XPS-INT-1}
\end{align}
 where $E_{0}$ and $E_{g}$ represent the total energy of the final
state $\left|f\left(0\right)\right\rangle $ and the ground state
$\left|g\right\rangle $ , $A$ is an energy independent constant
including the contribution of overlaps between the wave functions
of the core-state in the final and ground states. $N_{\mathrm{e}}$
is the number of valence electrons in the ground state. The overlaps
between the valence states and the core states are eliminated because
they could be almost orthogonal. $\Delta E_{\left(\ell,m\right)}$
is the excitation energy defined by $\Delta E_{\left(\ell,m\right)}=E_{\left(\ell,m\right)}-E_{0}=\sum_{\left(n,n^{\prime}\right)}(\epsilon_{n}-\epsilon_{n^{\prime}}),$
where $E_{\left(\ell,m\right)}$ represents the total energy of the
final state $\left|f\left(\ell,m\right)\right\rangle $ and $\epsilon_{n}$'s
are the Kohn-Sham eigenvalues. The energy difference $\epsilon_{n}-\epsilon_{n^{\prime}}$
stands for the energy of an e-h pair of an electron at level $n$
and a hole at level $n^{\prime}$ and the summation are taken over
all e-h pairs in the final state $\left|f\left(\ell,m\right)\right\rangle $.
Although the Kohn-Sham eigenvalues may not be proper quasi particle
energies, they practically give a good approximation to quasiparticle
energies, except for the fundamental energy gap.\cite{Hybertsen1986,Hamada1990}
In the following calculation, we replace the $\delta$ function by
the Lorentzian function with the full width at half maximum (FWHM)
$2\Gamma=1\,\mathrm{eV}$ in order to take into account the lifetime
broadening of the core level. $E_{0}-E_{G}$ is treated as an adjustable
parameter so that the threshold of XPS spectra coincides with the
experiments. In order to suppress the error caused by the fictitious
periodicity of the core-hole site, we pick up only the $\Gamma$ point
as the sample states for calculating XPS spectra. For Ni, we pick
up the X point $\left(\frac{1}{2},0,0\right)$ as a sample point in
addition to the $\Gamma$ point, because the $3d$ band states at
the $\Gamma$ point are fully occupied by both up- and down-spin electrons
on the system of the $3\times3\times3$ fcc super-cell. We take account
of the final states including $0$, $1$, $2$, and $3$ electron-hole
pairs on the final state $\left|f\left(0\right)\right\rangle $ and
restrict the final states $\left|f\left(\ell,m\right)\right\rangle $
so that the excitation energy $E_{\left(\ell,m\right)}-E_{0}$ is
less than $10$ eV. We need to calculate the more than $10^{8}$ determinants
of the matrices of the size of about $150\times150$ for Fe even after
the above simplification.

Before closing this subsection, we briefly mention the limitations
in this calculation. First, we assume that the $2p$ core hole is
spherically distributed with neglecting the dependence on the magnetic
quantum number of the core hole. Second, we take no account of the
spin-orbit interaction (SOI) in the band structure calculation. Third,
due to the finiteness of the cell size, the final state $|f\left(0\right)\rangle$
(no e-h pair) has a finite overlap with the ground state $|g\rangle$,
resulting in a finite intensity at the threshold. In principle, such
overlap should converge to zero with $N_{e}\to\infty$, according
to the Anderson orthogonality theorem;\cite{Anderson1967} energy
levels become continuous near the Fermi level and thereby infinite
numbers of e-h pairs could be created with infinitesimal excitation
energies, leading to the so-called Fermi edge singularity in the XPS
spectra. The finite contribution obtained above arises from the discreteness
of energy levels and could be interpreted as the integrated intensity
of singular spectra near the threshold, in consistent with the model
calculations for other systems.\cite{Kotani1974,Feldkamp1980}

\subsection{Binding energy difference at the threshold\label{sub:Binding-energy-difference}}

The threshold energies $\omega_{\uparrow}^{\mathrm{th}}$ for the
up-spin core-electron removal excitation and $\omega_{\downarrow}^{\mathrm{th}}$
for the down-spin one are generally different in the ferromagnetic
systems. It could be naively considered that the difference $\Delta\omega^{\mathrm{th}}=\omega_{\uparrow}^{\mathrm{th}}-\omega_{\downarrow}^{\mathrm{th}}$
is linked to the energy difference between the up- and down-spin core
levels in the ground state and is a good indicator for the local spin
moment. However, because of the considerable core-hole screening in
metals, the final state effects should be taken into consideration.
Within the fully screened potential approximation, we may obtain a
better estimate of $\Delta\omega^{\mathrm{th}}$ without using supercell
but by using the CPA in the low concentration limit, as was done in
the calculation of the core-level chemical shift in metallic random
alloys.\cite{Olovsson.PRL.87.176403,Olovsson.PRB.72.064203} Because
the core-hole screening is almost completed within the core-hole site
for metals, the CPA, which is a one-site approximation, may not cause
large error.

The threshold energy may be written as 
\begin{align}
\omega_{\uparrow\left(\downarrow\right)}^{\mathrm{th}} & =E_{\uparrow\left(\downarrow\right)}-\varepsilon_{\mathrm{F}}-E_{g},\label{eq:Threshold}
\end{align}
 where $E_{g}$ represents the total energy of the ground state, $E_{\uparrow\left(\downarrow\right)}$
does that of the fully screened state with an up(down)-spin core hole
and $N_{\mathrm{e}}+1$ band electrons, and $\varepsilon_{\mathrm{F}}$
is the Fermi energy in the fully screened state. Therefore, the difference
may be given by 
\begin{align}
\Delta\omega^{\mathrm{th}} & =\omega_{\uparrow}^{\mathrm{th}}-\omega_{\downarrow}^{\mathrm{th}}=\left(E_{\uparrow}-E_{g}\right)-\left(E_{\downarrow}-E_{g}\right).\label{eq:Diff Threshold}
\end{align}
 The energy difference $E_{\uparrow\left(\downarrow\right)}-E_{g}$
corresponds to the energy of putting one core-ionized impurity at
a normal site. This might be given as the generalized thermodynamic
chemical potential, 
\begin{equation}
E_{\uparrow\left(\downarrow\right)}-E_{g}=\left.\frac{\partial\overline{E}_{\uparrow\left(\downarrow\right)}\left(\rho\right)}{\partial\rho}\right|_{\rho\rightarrow0},\label{eq:GTCP}
\end{equation}
 where $\overline{E}_{\uparrow\left(\downarrow\right)}$ represents
the total energy per unit cell for the random alloy system which consist
of core-ionized and normal atoms and $\rho$ the density of the cell
including a core-hole site. Hence, the difference $\Delta\omega^{\mathrm{th}}$
may be written as 
\begin{align}
\Delta\omega^{\mathrm{th}} & =\left.\frac{d}{d\rho}\left(\overline{E}_{\uparrow}\left(\rho\right)-\overline{E}_{\downarrow}\left(\rho\right)\right)\right|_{\rho\rightarrow0}.\label{eq:Diff Threshold 2}
\end{align}
 We evaluate these values with the help of the KKR Green's function
band structure calculation combined with CPA.\cite{Akai1982,Akai1998}
We carry out the calculation at the concentration ranged from $\rho=0$
to $0.05$. $\Delta\omega^{\mathrm{th}}$ is calculated by an extrapolation
to zero concentration $\rho$.

\section{Results and Discussions}

\subsection{Ground and fully screened states}

\subsubsection{$d$-DOS}

We carry out the band structure calculation based on the LDA. For
the $2p$ electron removal, we make the FLAPW super cell calculation.
The magnetic moments in the ground states are obtained $2.1$, $1.6$,
and $0.6$ $\mathrm{\mu_{B}}$ inside a muffin-tin sphere ($r_{\mathrm{m}}=2.0\,\mathrm{Bohr}$)
for Fe, Co, and Ni, respectively. The obtained electronic structures
in the ground states are consistent with the past band structure calculations.\cite{Morruzi1978}
We have also checked the accuracy of the calculation by carrying out
the KKR band structure calculation combined with the CPA.\cite{Akai1982,Akai1998}
We treat the system as an random alloy system consisting of the normal
and core-ionized atoms, and take the limit of zero concentration of
the core-ionized atom. The FLAPW supercell calculation and the KKR-CPA
calculation give essentially the same results.

\begin{figure*}
\hfill{}\includegraphics[angle=-90,scale=0.6]{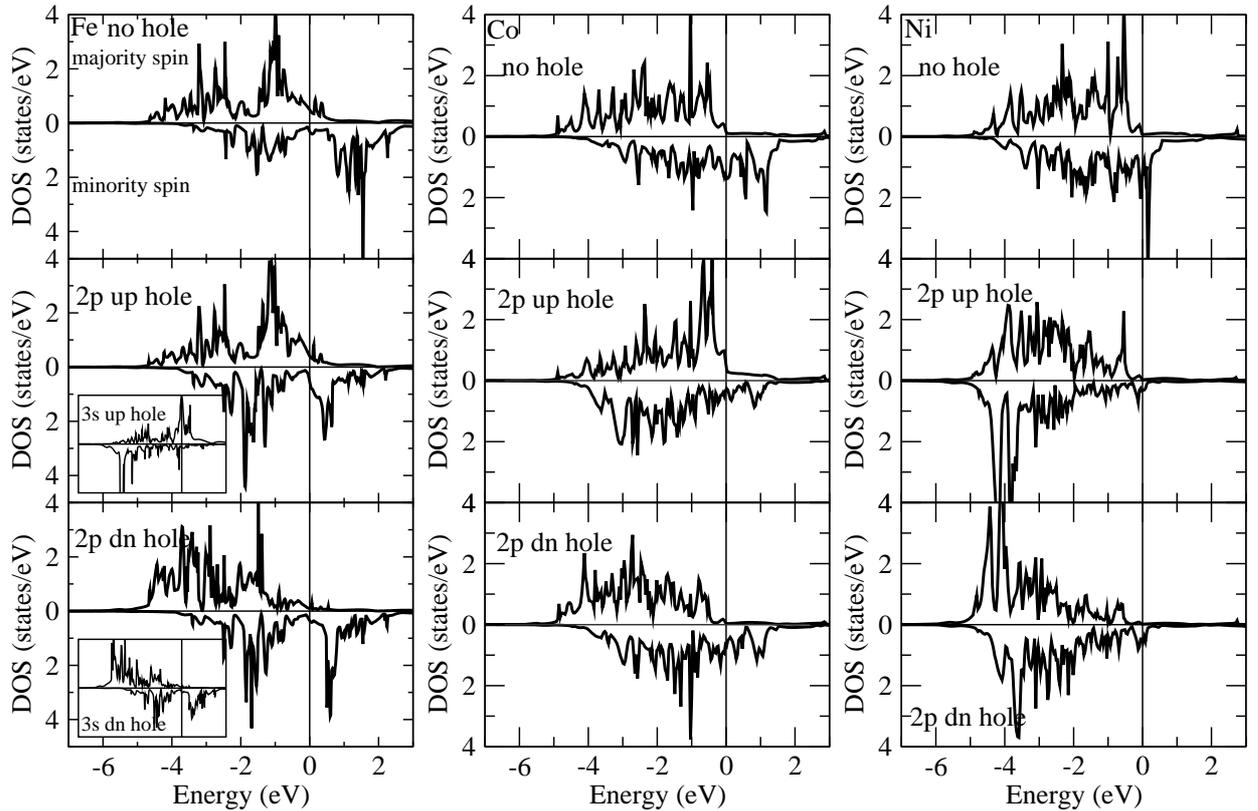}\hfill{}

\caption{Density of state projected onto the state with $d$ symmetry inside
the muffin-tin sphere ($d$-DOS) at the core-hole site in Fe, Co,
and Ni. Top panels show the $d$-DOS's in the ground states. Middle
(bottom) panels show those in the fully screened states with an up
(down)-spin $2p$ core-hole. The $d$-DOS's in the fully screened
states with a $3s$ hole are also shown as inset. The energy of the
Fermi level is zero. \label{fig:Fe-Co-Ni-2p-core-hole}}
\end{figure*}

The one-electron states in the fully screened state are modified from
those in the ground state by the core-hole potential. Such changes
in one-electron states may be explained through the change in the
$d$-symmetric density state ($d$-DOS) at the core-hole site. Figure
\ref{fig:Fe-Co-Ni-2p-core-hole} shows the $d$-DOS at the core-hole
site. The $d$-DOS's at the core-hole site in the fully screened states
are greatly modified from those in the ground state. The $d$-DOS
at sites with no core hole, which are not shown here, are almost unchanged
from those in the ground state. This indicates that the core-hole
potential is almost completely screened inside the core-hole site,
being consistent with the small screening length $\sim$Bohr radius
for metals. The change of the $d$-DOS's by the core-hole potential
depends largely on whether the up- or down-spin electron is removed
from the $2p^{6}$ core state. We refer to the hole created in the
$2p$ states by removing an up(down)-spin electron as the up(down)-spin
hole.

In the fully screened state with an up-spin $2p$ hole, the weight
of the down-spin $d$-DOS moves toward the higher binding energy region
if compared with the up-spin $d$-DOS. This is due to the repulsive
(attractive) exchange interaction working between the up-spin hole
and the up(down)-spin $3d$ electrons, in addition to the attractive
interaction between the core hole and $3d$ electrons. In Ni, the
weight of the $d$-DOS of the down-spin state concentrate near the
bottom of the $3d$ band, indicative of the quasibound states. The
weight of the up-spin states also shifts toward to the higher binding
energy region. These behaviors are very similar to those for the $3s$
electron removal. In Co, the weight of the $d$-DOS's of the down-spin
$3d$ states shifts toward the higher binding energy region. Opposite
to the case for Ni, that weight of the up-spin $3d$ states shifts
toward the Fermi level. The behaviors are similar to those for the
$3s$ electron removal but with rather moderate shape. In Fe, the
weight of the $d$-DOS's of the down-spin $3d$ state moves toward
the higher binding energy region, but not as much as for Ni. The weight
of the up-spin $3d$ states is almost unchanged. The behavior in Fe
contrasts to that for the $3s$ electron removal, where the weight
of the down-spin $d$-DOS highly concentrates near the bottom of the
$3d$ band and that for the up-spin $d$-DOS significantly shift toward
the Fermi level.

In the fully screened state with a down-spin $2p$ hole, the weight
of the up-spin $d$-DOS's tends to move toward the higher binding
energy region probably due to the exchange interaction between the
core hole and the $3d$ electrons. In Fe, the weight of the up $d$-DOS
is moderately shifted toward the higher binding energy region compared
to the case of the down-spin $3s$ electron removal. The weight of
the down-spin $d$-DOS also slightly moves toward the higher binding
energy region. In Ni, the shift of the up-spin $d$-DOS is large,
indicative of the quasibound states. The weight of the down-spin $d$-DOS
also concentrates near the bottom of the $3d$ band. This behavior
in Ni is very similar to that for the down-spin $3s$ electron removal.
The behavior of the $d$-DOS's in Co is intermediate between Fe and
Ni. To sum up, the $3d$ states in Fe for the $2p$ electron removal
are quite different from those for the $3s$ electron removal, while
those in Ni are quite similar in both removal states.

\subsubsection{Screening electron number}

The change of the $d$-DOS's is related to the screening electron
number at the core-hole site. Table \ref{tab:Screening-electron-number}
shows the screening electron numbers $\Delta n_{d\uparrow}$, $\Delta n_{d\downarrow}$,
and $\Delta n_{d\uparrow}+\Delta n_{d\downarrow}$. Here $\Delta n_{d\uparrow\left(\downarrow\right)}$
is given by $\Delta n_{d\uparrow\left(\downarrow\right)}=n_{d\uparrow\left(\downarrow\right)}^{\mathrm{sc}}-n_{d\uparrow\left(\downarrow\right)}^{\mathrm{gr}}$,
where $n_{d\uparrow\left(\downarrow\right)}^{\mathrm{sc}}$ is the
up(down)-spin electron number inside the muffin-tin sphere in the
$d$-symmetric states in the core-hole site in the fully screened
state, and $n_{d\uparrow\left(\downarrow\right)}^{\mathrm{gr}}$ is
the corresponding quantity in the ground state. Total screening electron
numbers $\Delta n_{d\uparrow}+\Delta n_{d\downarrow}$ inside the
muffin-tin sphere is found to be roughly unity. On the sites with
no core hole, the electron numbers are almost unchanged from that
in the ground state, indicating that the core-hole screening is almost
completed inside the core-hole site.

First we discuss the case that an up-spin $2p$ electron is removed.
In the fully screened state of Co, the core hole attracts so strongly
the down-spin $3d$ electrons that they overscreen the core hole,
and, as a countereffect, a few up-spin $3d$ electrons at the core-hole
site are pushed away from the muffin-tin sphere due to the Coulomb
repulsion with the excess down-spin electrons. In the fully screened
state of Ni, although the strong attraction by the core hole is expected
to work on the down-spin electrons, the screening down-spin electron
number is less than unity, because the $d$ states at the core-hole
site are already almost filled in the ground state. The up-spin $3d$
electrons participate a little to screen the core-hole. Note that
those screening electron numbers are similar to those for the up-spin
$3s$ electron removal state both in Co and Ni. In Fe, the screening
of the core hole is almost completed by the down-spin $3d$ electrons
without overscreening. This is quite different from the fully screened
state for the up-spin $3s$ electron removal, where the down-spin
$3d$ electrons extremely overscreen the core hole, and that overscreening
is compensated by pushing away the up-spin $3d$ electrons from the
core hole site. Now we discuss the case that a down-spin $2p$ electron
is removed. In the fully screened states of Ni, although the up-spin
$3d$ electrons are to be strongly attracted by the core hole, the
screening electron number is not large, because the up-spin $3d$
state is almost filled in the ground state. Therefore, the down-spin
$3d$ electrons contribute in large amount, leading that the total
screening numbers become nearly unity. In the fully screened states
of Fe and Co, although the up-spin $3d$ electrons are more attracted
by the core hole than the down-spin electrons, the screening electron
number of the up-spin electron is smaller than the down-spin electrons
for a similar reason to Ni.

\begin{table}
\caption{Screening electron number with the $d$ symmetry inside the muffin-tin
sphere at the $2p$ core hole site. The radii of the muffin-tin spheres
are $2.0$ Bohr. \label{tab:Screening-electron-number}}

\label{table.1-1}

\hfill{}%
\begin{tabular}{rrlrrrrrc}
\hline 
 &  & $2p$ hole  &  & $\Delta n_{d\uparrow}$  &  & $\Delta n_{d\downarrow}$  &  & $\Delta n_{d\uparrow}+\Delta n_{d\downarrow}$ \tabularnewline
\hline 
Fe  &  & Up  &  & $0.19$  &  & $1.01$  &  & $1.20$\tabularnewline
 &  & Down  &  & $0.47$  &  & $0.67$  &  & $1.14$\tabularnewline
Co  &  & Up  &  & $-0.14$  &  & $1.35$  &  & $1.21$\tabularnewline
 &  & Down  &  & $0.22$  &  & $0.86$  &  & $1.08$\tabularnewline
Ni  &  & Up  &  & $0.18$  &  & $0.89$  &  & $1.07$\tabularnewline
 &  & Down  &  & $0.27$  &  & $0.79$  &  & $1.06$\tabularnewline
\hline 
\end{tabular}\hfill{}
\end{table}

\subsubsection{Difference between the $2p$ core electron removal and the $3s$-core
electron removal}

Now we focus on the difference in the screening behavior between with
the $2p$ core hole and the $3s$ core hole. For both cases, the down(up)-spin
$3d$ electrons are, generally speaking, attracted to the core hole
more strongly than the up(down)-spin $3d$ electrons in the fully
screened state both with an up(down)-spin $2p$ or $3s$ core hole,
because the exchange interaction is working between the core hole
and the $3d$ electrons. In addition to this tendency, the screening
in Fe seems moderate in the fully screened state with a $2p$ core
hole, compared to the presence of the overscreening with a $3s$ core
hole. On the other hand, in Ni, the screening with a $2p$ core hole
is comparably larger than that with a $3s$ core hole. How does such
a different screening behavior occur in Fe and Ni? The exchange interaction
$J_{2p\mbox{-}3d}$ between the $2p$ and $3d$ electrons, and $J_{3s\mbox{-}3d}$
between the $3s$ and $3d$ electron may be estimated by the atomic
Hartree-Fock calculation, which are $J_{2p\mbox{-}3d}\vphantom{=\frac{2}{15}G^{1}\left(2p,3d\right)+\frac{3}{35}G^{3}\left(3s,3d\right)}\sim1$
eV and $J_{3s\mbox{-}3d}\vphantom{=\frac{1}{5}G^{2}\left(3s,3d\right)}\sim2$
eV, for all the neutral Fe, Co, and Ni atoms, respectively. The small
$2p$-$3d$ exchange interaction in Fe could not solve the above question.

\begin{figure*}
\hfill{}\includegraphics[angle=-90,scale=0.6]{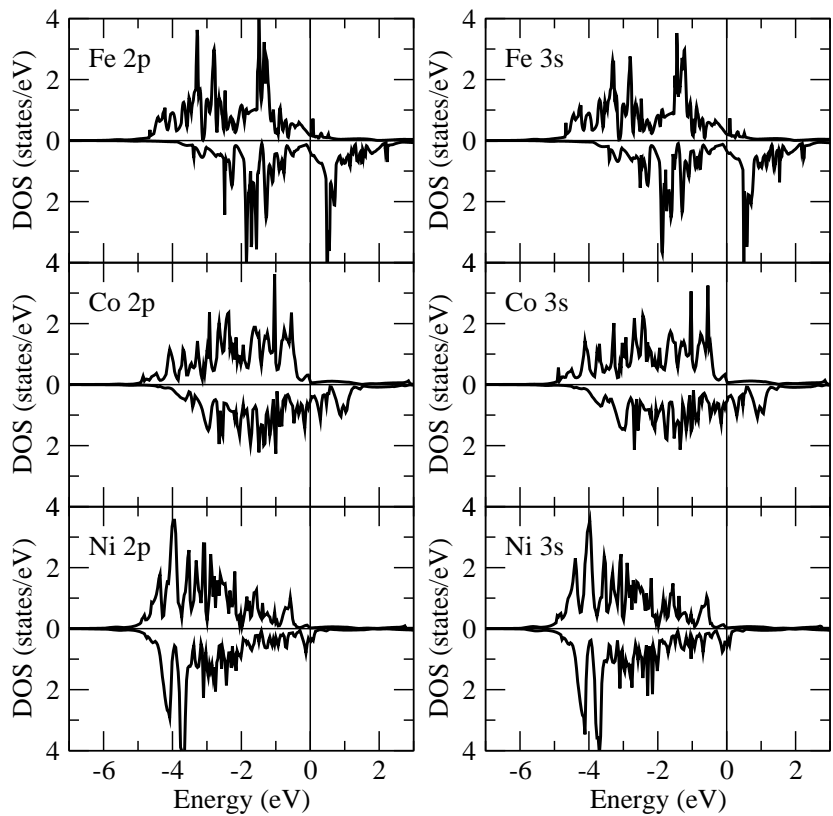}\hfill{}

\caption{$d$-DOS's when removing a half up- and a half down-spin electron
from the $2p$ and $3s$ core. The energy of the Fermi level is zero.
\label{fig:0-pol_dDOS}}
\end{figure*}

To eliminate the effect of the $2p$-$3d$ ($3s$-$3d$) exchange
interaction, we carry out the band calculation under the condition
that a half up-spin and a half down-spin electron is removed from
the $2p$ ($3s$) core states. Figure \ref{fig:0-pol_dDOS} shows
the $d$-DOS's at the core hole site in the fully screened state.
The $d$-DOS's for the state with a $2p$ core hole are similar to
those with a $3s$ core hole. Note that the atomic Hartree-Fock calculation
gives quite different values of the Slater integrals $F^{0}$'s for
$2p$-$3d$ and $3s$-$3d$ interactions, that is, $F^{0}\left(2p,3d\right)-F^{0}\left(3s,3d\right)\sim7\,\mathrm{eV}$,
which in fact contradicts the results shown in Fig.~\ref{fig:0-pol_dDOS}.
In Ni, the $d$-DOS's concentrate near the bottom of the up- and down-spin
$3d$ band, forming quasibound states. In Fe, the situation is quite
different; the weights of the $d$-DOS shift only moderately toward
the higher binding energy region, almost keeping the structures of
the ground state. In Co, the behavior of the $d$-DOS is intermediate
between Fe and Ni. This difference in Fe, Co, and Ni may be related
to the different electron occupation in the band in the ground state.
In Ni, because the $3d$ band is almost fully occupied, it is hard
to make enough room for accommodating the screening electron at the
core hole site by mixing the one-electron states within the $3d$
band states. Therefore, the mixing beyond the $3d$ band states has
to take place to complete core hole screening inside the core hole
site. These processes might help to form quasibound states. In Fe,
because the $3d$ band in the ground state has enough room to accommodate
excess screening electron, the mixing of the one-electron states within
the $3d$ band states might be sufficient in order to complete core
hole screening, resulting in only the slight shift of the weight of
the $d$-DOS's. When the $2p$-$3d$ ($3s$-$3d$) exchange potential
is turned on, the screened states get modified further. In Ni, both
the $2p$-$3d$ and $3s$-$3d$ exchange potentials simply shift the
quasibound states to the higher or lower binding energy regions. In
Fe, the $3s$-$3d$ exchange potential considerably modify the $3d$
state by pulling down the down-spin $3d$ states at around $0.5$
eV above the Fermi level (the top panel in Fig. \ref{fig:0-pol_dDOS})
under the Fermi level, and by making the quasibound states formed
near the bottom of the down-spin $3d$ band. The $2p$-$3d$ exchange
potential, on the other hand, hardly affects the $3d$ band states,
that is, the down-spin $3d$ states at around $0.5$ eV still stay
above the Fermi level probably due to the smallness of the $2p$-$3d$
exchange potential.

\subsection{Photoemission Spectra}

\begin{figure}
\hfill{}\includegraphics[scale=0.5]{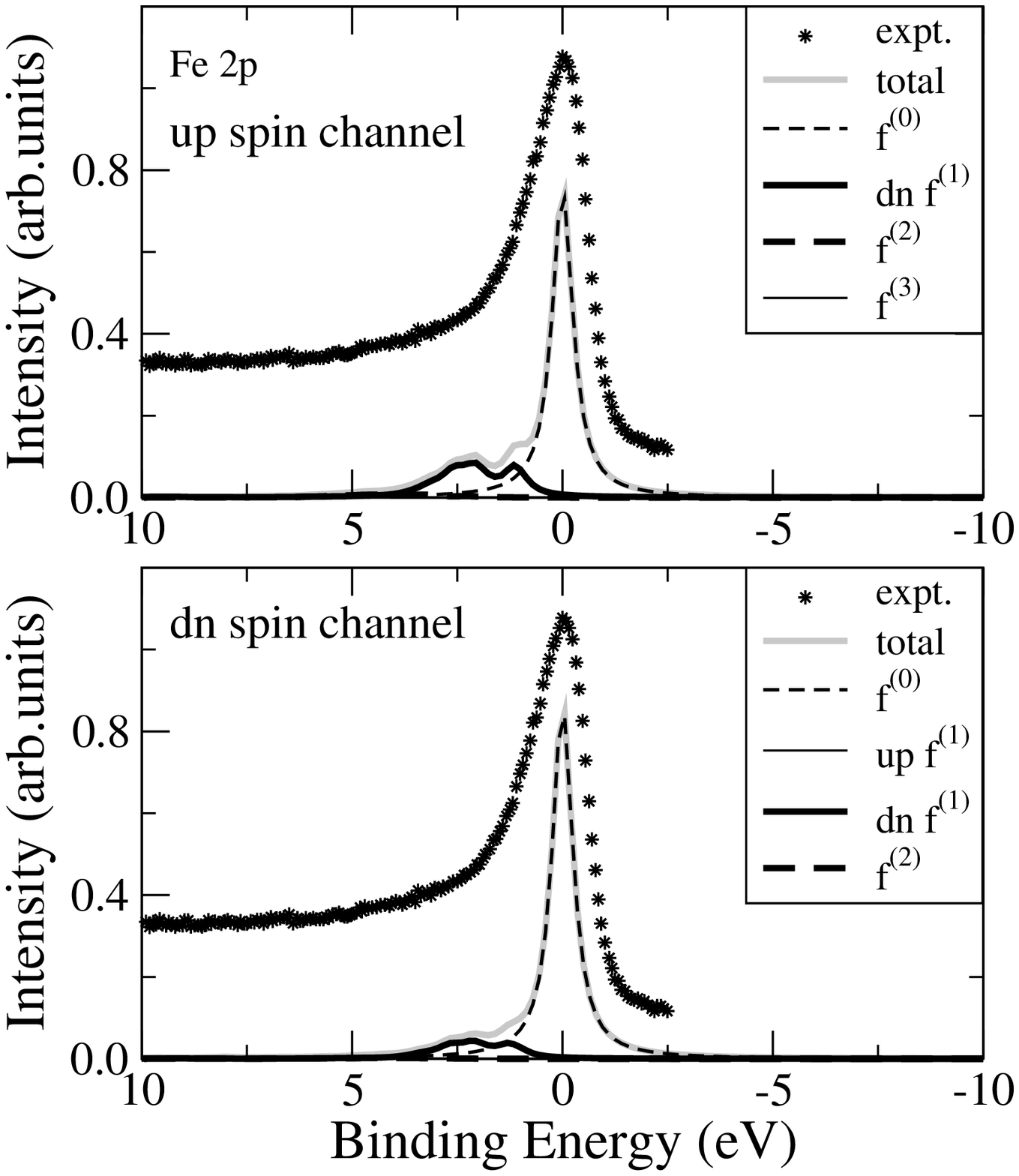}\hfill{}

\caption{Fe 2p XPS spectra. Gray thick curve represents the total intensity.
The curves denoted $f^{\left(0\right)}$, up $f^{\left(1\right)}$,
dn $f^{\left(1\right)}$, $f^{\left(2\right)}$, and $f^{\left(3\right)}$
are the contribution from the final states including zero, one up-spin
e-h pair, one down-spin e-h pair, two e-h pairs, and three e-h pairs
on the final state $\left|f\left(0\right)\right\rangle $, respectively.
Experimental data are taken from Ref. \onlinecite{Acker1988}, which
are not spin resolved.\label{fig:Fe-2p-XPS}}
\end{figure}

\begin{figure}
\hfill{}\includegraphics[scale=0.5]{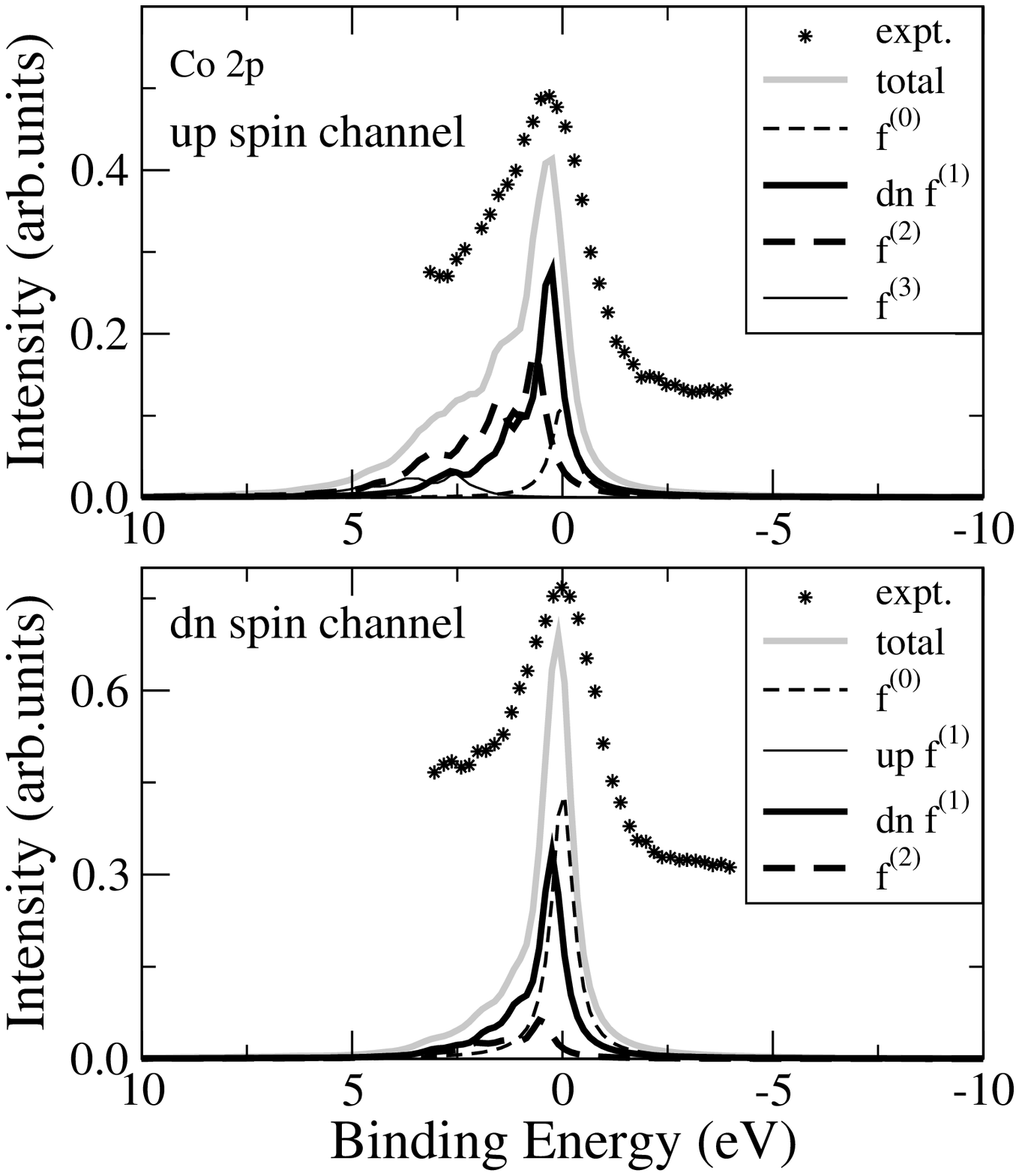}\hfill{}

\caption{Co 2p XPS spectra.Gray thick curve represents the total intensity.
The curves denoted $f^{\left(0\right)}$, up $f^{\left(1\right)}$,
dn $f^{\left(1\right)}$, $f^{\left(2\right)}$, and $f^{\left(3\right)}$
are the contribution from the final states including zero, one up-spin
e-h pair, one down-spin e-h pair, two e-h pairs, and three e-h pairs
on the final state $\left|f\left(0\right)\right\rangle $, respectively.
Spin-resolved experimental data are taken from Ref. \onlinecite{Klebanoff1994}.\label{fig:Co-2p-XPS}}
\end{figure}

\begin{figure}
\hfill{}\includegraphics[scale=0.5]{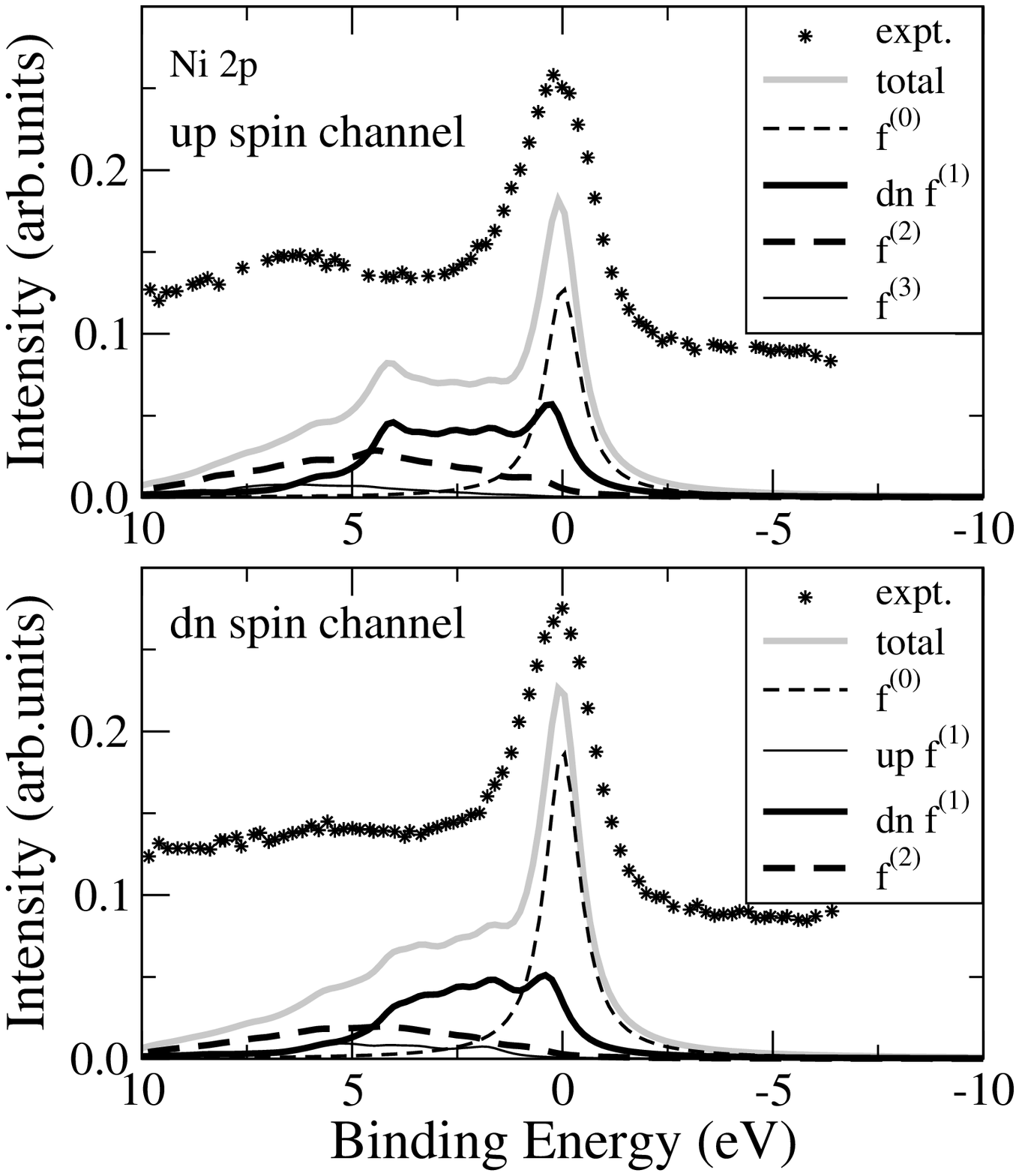}\hfill{}

\caption{Ni 2p XPS spectra. Gray thick curve represents the total intensity.
The curves denoted $f^{\left(0\right)}$, up $f^{\left(1\right)}$,
dn $f^{\left(1\right)}$, $f^{\left(2\right)}$, and $f^{\left(3\right)}$
are the contribution from the final states including zero, one up-spin
e-h pair, one down-spin e-h pair, two e-h pairs, and three e-h pairs
on the final state $\left|f\left(0\right)\right\rangle $, respectively.
Spin-resolved experimental data are taken from Ref. \onlinecite{See1995Ni}.\label{fig:Ni-2p-XPS}}
\end{figure}

Using the one-electron wave functions in the ground state and fully
screened states, the XPS intensities are calculated as a function
of the binding energy from eq. (\ref{eq:XPS-INT-1}). Figures \ref{fig:Fe-2p-XPS}--\ref{fig:Ni-2p-XPS}
show the calculated spectra for bcc Fe, fcc Co, and fcc Ni, in comparison
with the experimental $2p_{3/2}$ spectra. Although the edge singularity
is not reproduced, the overlaps $\left\langle f\left(0\right)|g\right\rangle $
and $\left\langle f\left(1,m\right)|g\right\rangle $'s appear to
give the reasonable intensities around threshold. The total spectral
curves, which depend on the elements, reproduce well the overall structures
observed in the experiments. Note that the final states $\left|f\left(1,m\right)\right\rangle $'s
holding a down-spin e-h pair mainly contribute to the intensity. The
final state $\left|f\left(1,m\right)\right\rangle $'s created by
putting an up e-h pair on the final state $\left|f\left(0\right)\right\rangle $
hardly contribute to the intensity, because the up-spin $3d$ band
states are almost filled in the ground state, and thereby the overlap
determinant become quite small.

The Fe spectra show a single-peak structure with a weak shoulder intensity
for both the up- and down-spin $2p$ electron removals, contrasting
with the strong satellite intensity for the up-spin $3s$ electron
removal. Because the one-electron wave functions in the fully screened
state are not so strongly modified from those in the ground state,
the overlap $\left\langle f\left(0\right)|g\right\rangle $ is almost
unity. Consequently, the other overlaps $\left\langle f\left(\ell,m\right)|g\right\rangle $'s
are almost zero, because the one-electron wave functions for the unoccupied
states in the final state $\left|f\left(0\right)\right\rangle $ are
nearly orthogonal to those for occupied states in the ground state
$\left|g\right\rangle $. Note that the observed $0.9$ eV splitting
of the peak position around the Fe $2p_{3/2}$ threshold is not reproduced
by the e-h excitations on the final state $\left|f\left(0\right)\right\rangle $.
We expect that the splitting originates from the different threshold
energy for the up- and down-spin $2p$ electron removals.

In Co, the intensities at the higher binding energy region are larger
than the Fe $2p$ spectra, forming a rather strong shoulderlike structure
for the up-spin $2p$ electron removal. Because the one-electron wave
functions for the down-spin $3d$ band states in the fully screened
state are noticeably modified from those in the ground state, the
one-electron wave functions of the unoccupied levels in the final
state $\left|f\left(0\right)\right\rangle $ have some amplitudes
of those of the occupied levels in the ground state. The overlaps
$\left\langle f\left(\ell,m\right)|g\right\rangle $'s accordingly
become finite, giving rise to the shoulder. For the down-spin $2p$
electron removal, the spectra show a single-peak structure, because
the down-spin $3d$ states are not strongly modified by the down-spin
$2p$ hole.

The Ni spectra exhibit satellite structure for the up-spin $2p$ electron
removal. The e-h pair excitations that an electron on the quasibound
state is excited to the unoccupied states give rise to the satellite
intensity. The satellite peak position is around $4$ eV, which is
$2$ eV smaller than the so-called $6$-eV satellite observed in experiment.
This discrepancy might be caused by the LDA or the fully screened
potential approximation. Because both the up- and down-spin $3d$
states are fully occupied in the fully screened state, the $3d$-$3d$
Coulomb interaction may not be relevant to this discrepancy within
the fully screened potential approximation. Braicovich and van der
Laan\cite{Braicovich.PRB.78.174421} estimated the screening time
constant $\tau_{s}=1.5\,\mathrm{fs}$ in Ni, which is definitely longer
than the values $0.18\,\mathrm{fs}$ in Fe and $0.43\,\mathrm{fs}$
in Co. Because the constant $\tau_{s}$ is comparable to the core
hole lifetime $\tau_{c}\sim1\,\mathrm{fs}$, the use of the fully
screened potential may not be appropriate to describe the satellite
intensities in Ni. The spectra for the down-spin electron removal
also show relatively large shoulder intensities, which would come
from the excitations in the down-spin $3d$ states, because the unoccupied
levels are available in the down-spin $3d$ states in the fully screened
$3d$ states.

\subsection{Energy difference of the threshold}

\begin{figure}
\hfill{}\includegraphics[scale=0.6]{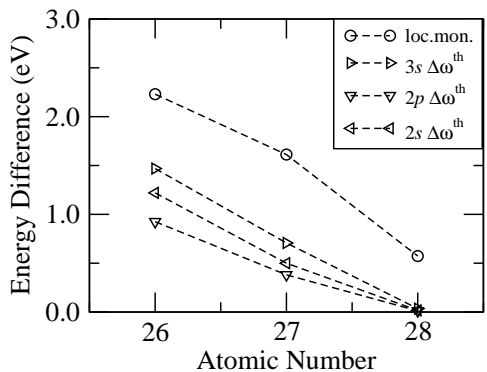}\hfill{}

\caption{Difference of the threshold energy between the states that an up-spin
core-electron is removed and that a down-spin core-electron is removed.
Left, down, and right triangles represent $\Delta\omega^{\mathrm{th}}$
for the $2s$, $2p$, and $3s$ core-ionized states. Circle represents
the local spin moment at the core hole site in $\mu_{\mathrm{B}}$.
Dashed lines are for a guide to eyes.\label{fig:Domega-spin}}
\end{figure}

\begin{figure}
\hfill{}\includegraphics[scale=0.6]{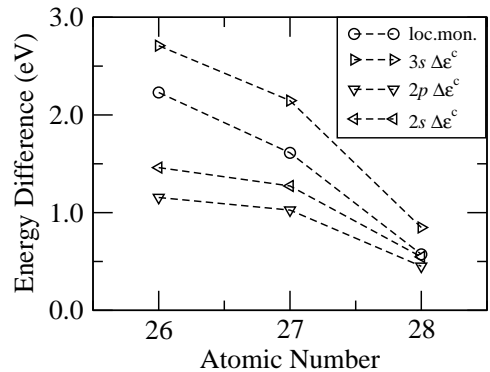}\hfill{}

\caption{Difference of the up-spin and the down-spin core-level energies. Left,
down, and right triangles represent $\Delta\varepsilon^{\mathrm{c}}$
for $2s$, $2p$, and $3s$ core-ionized states. Circle represents
the local spin moment at the core hole site in $\mu_{\mathrm{B}}$.
Dashed lines are for a guide to eyes. \label{fig:Depsilon-spin}}
\end{figure}

In order to estimate the magnetic splitting of the threshold energy,
we calculate the energy difference of the threshold in the up- and
down-spin core-electron removals by using the KKR-CPA method, as discussed
in Sec.\ref{sub:Binding-energy-difference}. Figures \ref{fig:Domega-spin}
and \ref{fig:Depsilon-spin} show the energy difference of the threshold
$\Delta\omega^{\mathrm{th}}$ defined by Eq. (\ref{eq:Diff Threshold 2})
and the energy difference of the core levels $\Delta\varepsilon^{\mathrm{c}}=\varepsilon_{\downarrow}^{\mathrm{c}}-\varepsilon_{\uparrow}^{\mathrm{c}}$
in the ground states. While $\Delta\varepsilon^{\mathrm{c}}$'s are
nearly proportional to the local spin moment, $\Delta\omega^{\mathrm{th}}$'s
decrease more rapidly than the change of the local moments with moving
from Fe to Ni. In Fe, we get the local moment about $2.2\,\mathrm{\mu_{B}}$
and $\Delta\omega^{\mathrm{th}}$'s about $1.2$, $0.9$, and $1.5\,\mathrm{eV}$
for the $2s$, $2p$, and $3s$ excitations, respectively. In Ni,
on the other hand, $\Delta\omega^{\mathrm{th}}$'s are almost zero,
despite the definite local spin moment about $0.6\,\mathrm{\mu_{B}}$.
These values seem to be consistent with the experimental observations;
the splitting is estimated as $\gtrsim1\,\mathrm{eV}$ in the Fe $3s$
spin resolved spectra,\cite{Acker1988,Campen1994,Imada.unpub} and
$0.9$ eV in the Fe $2p_{3/2}$ HAXPES spectra.\cite{Imada.unpub}
On the other hand, such splittings have not been observed in the Ni
$2p$ and $3s$ spectra.

It is obvious that the core hole screening plays crucial roles in
determining the magnitude of the splitting of the threshold energy.
The fact that the magnetic splitting in Ni is smaller than that in
Fe may be understood as follows. All the $3d$ states in the fully
screened state in Ni are are pulled down below the Fermi level forming
the quasibound states resulting in a suppression of the local spin
moment at the core hole site solely by the Coulomb interaction between
the core hole and the $3d$ electrons (Fig. \ref{fig:0-pol_dDOS}).
The exchange potential could not give rise to a large difference in
the screening electron distribution between for the up-spin and down-spin
electron removals. The $3d$ states in Fe, on the other hand, are
strongly affected by the exchange potential particularly for the $3s$
electron removal, with the considerable change of the $d$-DOS's and
with the screening electron numbers. Even for the $2p$ electron removal,
the screening electron numbers depend considerably on the spin channel.
Thus we may guess that the weak effect of the exchange potential on
the $3d$ states in Ni may be the origin of the magnetic splitting
of the threshold energy. Note that the zero magnetic splitting does
not necessarily mean that the XPS spectra for the up-spin channel
are the same as those for the down-spin channel.

\section{Concluding Remarks}

We have applied an \emph{ab initio} method to calculate the $2p$
core-level XPS spectra as a function of binding energy in ferromagnetic
metals Fe, Co, and Ni. The calculated spectra have been compared to
the spin-resolved $2p_{3/2}$ spectra, where the up- and down-spin
electron removal excitations are considered separately. Because the
SOI is not included in the calculation, we could not distinguish $2p_{3/2}$
and $2p_{1/2}$. The dependence of the spectral shapes on the element,
the excited core, and the spin of the core hole left behind are well
reproduced by the calculation. The Fe $2p$ spectra show almost a
single-peak structure for the up-spin and down-spin electron removals.
In contrast, the spectral intensities for Ni are distributed in a
wide range of binding energy with the notable satellite structure
for the up-spin electron removal and the shoulder-like structure for
the down-spin electron removal. The spectra for Co exhibit intermediate
features between those for Fe and Ni; the spectra are widely distributed
around the higher binding energy region with noticeable shoulder structure
for the up-spin electron removal, while the spectra exhibit almost
a single-peak structure for the down-spin electron removal. The satellite
intensities in Ni $2p$ spectra are interpreted as coming from the
final state that an electron on the quasibound state is excited to
the unoccupied one-electron states on $\left|f\left(0\right)\right\rangle $.
Such satellite or shoulder structures exist in all elements Fe, Co,
and Ni for the up-spin $3s$ electron removal.\cite{Taka2010FeCoNi3sXPS}
We have discussed the difference between the $2p$ and $3s$ spectra
as well as the difference between Fe and Ni spectra in connection
with the one-electron states screening the core hole. We have explained
the origin of these differences by a combined effect of the different
magnitude of the $2p$-$3d$ and $3s$-$3d$ exchange potentials and
the different occupation numbers in the $3d$ states in Fe and Ni.

Although the calculation reproduce consistently the spectra shape
depending on the elements, the excited core, and the spin of the core
hole left behind, there is a clear discrepancy on the satellite position
in Ni; $6\,\mathrm{eV}$ satellite in Ni that the calculation gives
the satellite is given at merely $4\,\mathrm{eV}$ higher than the
threshold. One reason for this discrepancy may be that the fully screened
potential approximation is not appropriate due to the insufficient
time of screening core hole in Ni in comparison with the core hole
lifetime. Another reason might be caused by the approximate nature
of LDA. To clarify this issue, we need further studies.

In connection with the magnetic splittings of the threshold energy,
we have directly evaluated these values within the fully screened
potential approximation by using the KKR-CPA method. Olovsson \emph{et
al}. \cite{Olovsson.PRL.87.176403,Olovsson.PRB.72.064203} have discussed
the core-level chemical shifts in several metallic alloys using the
same method, and pointed out that the fully screened potential approximation
gives good correspondence to the experiments. The core-level energy
depends on its spin, which difference is roughly proportional to the
local spin moment. We have found that the magnetic splitting of the
threshold energy decreases more rapidly than the local spin moment
with moving from Fe to Ni; it is almost zero for Ni for the $2s$,
$2p$, and $3s$ electron removals, in spite of a finite local moment
$0.6\,\mu_{\mathrm{B}}$. The splitting may be directly measured by
the shift of the peak around the threshold in the spin resolved $2s$
or $3s$ spectra. Actually, in the recent experiment of the $2p_{3/2}$
HAXPES spectra, \cite{Imada.unpub} such a splitting is measured as
$0.9$ eV. Our calculation also suggests that the splitting would
hardly be observed in Ni. To be more precise, it may be better to
take account of the SOI and to increase the super-cell size.

We have considered only the one-electron states on the $\Gamma$ point
in the first Brillouin zone onto which electrons are distributed in
the calculation of the XPS spectra. This may not cause large errors
except the intensity near the threshold, because the first Brillouin
zone is reduced to a smaller size in a system of larger supercells.
With increasing the states onto which electrons are distributed, we
expect that the overlap between the lowest energy state in the presence
of core hole and the ground states would be reduced, and that the
contributions from e-h pair creation would increase near the threshold,
leading to an asymmetric peak near the threshold as a function of
binding energy. Such behavior has been demonstrated in numerical calculations
on finite-size systems.\cite{Kotani1974,Feldkamp1980} On the other
hand, the structures in the high binding energy region are expected
to be only a little influenced by such a refined treatment. In any
case, to be more quantitative, we need to increase the size of supercells
in the calculation.

Another \emph{ab initio} approach has been tried by using multiple
scattering theory.\cite{Kruger2004} Since the experimental data have
been accumulated for XPS spectra and the x-ray absorption spectra
near the $L$ edge, the extension of the present method to calculate
the absorption spectra is left in future study.
\begin{acknowledgments}
We used the FLAPW code developed by Noriaki Hamada, and the KKR-CPA
code by Hisazumi Akai. This work was partially supported by a Grant-in-Aid
for Scientific Research in Priority Areas (No.22540325) of The Ministry
of Education, Culture, Sports, Science, and Technology, Japan. 
\end{acknowledgments}
 \bibliographystyle{apsrev}
\bibliography{Bibcore}

\begin{thebibliography}{33}
\expandafter\ifx\csname natexlab\endcsname\relax\def\natexlab#1{#1}\fi
\expandafter\ifx\csname bibnamefont\endcsname\relax
  \def\bibnamefont#1{#1}\fi
\expandafter\ifx\csname bibfnamefont\endcsname\relax
  \def\bibfnamefont#1{#1}\fi
\expandafter\ifx\csname citenamefont\endcsname\relax
  \def\citenamefont#1{#1}\fi
\expandafter\ifx\csname url\endcsname\relax
  \def\url#1{\texttt{#1}}\fi
\expandafter\ifx\csname urlprefix\endcsname\relax\def\urlprefix{URL }\fi
\providecommand{\bibinfo}[2]{#2}
\providecommand{\eprint}[2][]{\url{#2}}

\bibitem[{\citenamefont{Anderson}(1967)}]{Anderson1967}
\bibinfo{author}{\bibfnamefont{P.~W.} \bibnamefont{Anderson}},
  \bibinfo{journal}{Phys.\ Rev.\ Lett.} \textbf{\bibinfo{volume}{18}},
  \bibinfo{pages}{1049} (\bibinfo{year}{1967}).

\bibitem[{\citenamefont{Mahan}(1967)}]{Mahan1967}
\bibinfo{author}{\bibfnamefont{G.~D.} \bibnamefont{Mahan}},
  \bibinfo{journal}{Phys.\ Rev.} \textbf{\bibinfo{volume}{163}},
  \bibinfo{pages}{612} (\bibinfo{year}{1967}).

\bibitem[{\citenamefont{Nozi\'eres and de~Dominicis}(1969)}]{Nozieres1969}
\bibinfo{author}{\bibfnamefont{P.}~\bibnamefont{Nozi\'eres}} \bibnamefont{and}
  \bibinfo{author}{\bibfnamefont{C.~T.} \bibnamefont{de~Dominicis}},
  \bibinfo{journal}{Phys.\ Rev.} \textbf{\bibinfo{volume}{178}},
  \bibinfo{pages}{1097} (\bibinfo{year}{1969}).

\bibitem[{\citenamefont{Doniach and Sunjic}(1970)}]{Doniach1970}
\bibinfo{author}{\bibfnamefont{S.}~\bibnamefont{Doniach}} \bibnamefont{and}
  \bibinfo{author}{\bibfnamefont{M.}~\bibnamefont{Sunjic}},
  \bibinfo{journal}{J.\ Phys. C} \textbf{\bibinfo{volume}{3}},
  \bibinfo{pages}{285} (\bibinfo{year}{1970}).

\bibitem[{\citenamefont{Hufner and Wetheim}(1975)}]{Hufner1975}
\bibinfo{author}{\bibfnamefont{S.}~\bibnamefont{Hufner}} \bibnamefont{and}
  \bibinfo{author}{\bibfnamefont{G.~W.} \bibnamefont{Wetheim}},
  \bibinfo{journal}{Phys.\ Lett.} \textbf{\bibinfo{volume}{51A}},
  \bibinfo{pages}{301} (\bibinfo{year}{1975}).

\bibitem[{\citenamefont{Feldkamp and Davis}(1980)}]{Feldkamp1980}
\bibinfo{author}{\bibfnamefont{L.~A.} \bibnamefont{Feldkamp}} \bibnamefont{and}
  \bibinfo{author}{\bibfnamefont{L.~C.} \bibnamefont{Davis}},
  \bibinfo{journal}{Phys.\ Rev.\ B} \textbf{\bibinfo{volume}{22}},
  \bibinfo{pages}{3644} (\bibinfo{year}{1980}).

\bibitem[{\citenamefont{van Acker et~al.}(1988)\citenamefont{van Acker,
  Stadnik, Fuggle, Hoekstra, Buschow, and Stroink}}]{Acker1988}
\bibinfo{author}{\bibfnamefont{J.~F.} \bibnamefont{van Acker}},
  \bibinfo{author}{\bibfnamefont{Z.~M.} \bibnamefont{Stadnik}},
  \bibinfo{author}{\bibfnamefont{J.~C.} \bibnamefont{Fuggle}},
  \bibinfo{author}{\bibfnamefont{H.~J. W.~M.} \bibnamefont{Hoekstra}},
  \bibinfo{author}{\bibfnamefont{K.~H.~J.} \bibnamefont{Buschow}},
  \bibnamefont{and} \bibinfo{author}{\bibfnamefont{G.}~\bibnamefont{Stroink}},
  \bibinfo{journal}{Phys. Rev. B} \textbf{\bibinfo{volume}{37}},
  \bibinfo{pages}{6827} (\bibinfo{year}{1988}).

\bibitem[{\citenamefont{Van~Campen and Klebanoff}(1994)}]{Campen1994}
\bibinfo{author}{\bibfnamefont{D.~G.} \bibnamefont{Van~Campen}}
  \bibnamefont{and} \bibinfo{author}{\bibfnamefont{L.~E.}
  \bibnamefont{Klebanoff}}, \bibinfo{journal}{Phys. Rev. B}
  \textbf{\bibinfo{volume}{49}}, \bibinfo{pages}{2040} (\bibinfo{year}{1994}).

\bibitem[{\citenamefont{Imada}(2011)}]{Imada.unpub}
\bibinfo{author}{\bibfnamefont{S.}~\bibnamefont{Imada}},
  \bibinfo{journal}{unpublished}  (\bibinfo{year}{2011}).

\bibitem[{\citenamefont{Plogmann et~al.}(1999)\citenamefont{Plogmann,
  Schlath\"olter, Braun, Neumann, Yarmoshenko, Yablonskikh, Shreder, Kurmaev,
  Wrona, and \ifmmode~\acute{S}\else \'{S}\fi{}lebarski}}]{Plogmann1999}
\bibinfo{author}{\bibfnamefont{S.}~\bibnamefont{Plogmann}},
  \bibinfo{author}{\bibfnamefont{T.}~\bibnamefont{Schlath\"olter}},
  \bibinfo{author}{\bibfnamefont{J.}~\bibnamefont{Braun}},
  \bibinfo{author}{\bibfnamefont{M.}~\bibnamefont{Neumann}},
  \bibinfo{author}{\bibfnamefont{Y.~M.} \bibnamefont{Yarmoshenko}},
  \bibinfo{author}{\bibfnamefont{M.~V.} \bibnamefont{Yablonskikh}},
  \bibinfo{author}{\bibfnamefont{E.~I.} \bibnamefont{Shreder}},
  \bibinfo{author}{\bibfnamefont{E.~Z.} \bibnamefont{Kurmaev}},
  \bibinfo{author}{\bibfnamefont{A.}~\bibnamefont{Wrona}}, \bibnamefont{and}
  \bibinfo{author}{\bibfnamefont{A.}~\bibnamefont{\ifmmode~\acute{S}\else
  \'{S}\fi{}lebarski}}, \bibinfo{journal}{Phys. Rev. B}
  \textbf{\bibinfo{volume}{60}}, \bibinfo{pages}{6428} (\bibinfo{year}{1999}).

\bibitem[{\citenamefont{\'Slebarski et~al.}(2001)\citenamefont{\'Slebarski,
  Nuemann, and Schneider}}]{Slebarski2001}
\bibinfo{author}{\bibfnamefont{A.}~\bibnamefont{\'Slebarski}},
  \bibinfo{author}{\bibfnamefont{M.}~\bibnamefont{Nuemann}}, \bibnamefont{and}
  \bibinfo{author}{\bibfnamefont{B.}~\bibnamefont{Schneider}},
  \bibinfo{journal}{J. Phys. : Condens. Matter} \textbf{\bibinfo{volume}{13}},
  \bibinfo{pages}{5515 } (\bibinfo{year}{2001}).

\bibitem[{\citenamefont{Shukla et~al.}(2007)\citenamefont{Shukla, Kr\"uger,
  Dhaka, Sayago, Horn, and Barman}}]{Shukla2007}
\bibinfo{author}{\bibfnamefont{A.~K.} \bibnamefont{Shukla}},
  \bibinfo{author}{\bibfnamefont{P.}~\bibnamefont{Kr\"uger}},
  \bibinfo{author}{\bibfnamefont{R.~S.} \bibnamefont{Dhaka}},
  \bibinfo{author}{\bibfnamefont{D.~I.} \bibnamefont{Sayago}},
  \bibinfo{author}{\bibfnamefont{K.}~\bibnamefont{Horn}}, \bibnamefont{and}
  \bibinfo{author}{\bibfnamefont{S.~R.} \bibnamefont{Barman}},
  \bibinfo{journal}{Phys. Rev. B} \textbf{\bibinfo{volume}{75}},
  \bibinfo{pages}{235419} (\bibinfo{year}{2007}).

\bibitem[{\citenamefont{Cui et~al.}(2008)\citenamefont{Cui, Kimura, Miyamoto,
  Taniguchi, Xie, Qiao, Shimada, Namatame, Ikenaga, Kobayashi
  et~al.}}]{Cui2008}
\bibinfo{author}{\bibfnamefont{Y.~T.} \bibnamefont{Cui}},
  \bibinfo{author}{\bibfnamefont{A.}~\bibnamefont{Kimura}},
  \bibinfo{author}{\bibfnamefont{K.}~\bibnamefont{Miyamoto}},
  \bibinfo{author}{\bibfnamefont{M.}~\bibnamefont{Taniguchi}},
  \bibinfo{author}{\bibfnamefont{T.}~\bibnamefont{Xie}},
  \bibinfo{author}{\bibfnamefont{S.}~\bibnamefont{Qiao}},
  \bibinfo{author}{\bibfnamefont{K.}~\bibnamefont{Shimada}},
  \bibinfo{author}{\bibfnamefont{H.}~\bibnamefont{Namatame}},
  \bibinfo{author}{\bibfnamefont{E.}~\bibnamefont{Ikenaga}},
  \bibinfo{author}{\bibfnamefont{K.}~\bibnamefont{Kobayashi}},
  \bibnamefont{et~al.}, \bibinfo{journal}{Phys. Rev. B}
  \textbf{\bibinfo{volume}{78}}, \bibinfo{pages}{205113}
  (\bibinfo{year}{2008}).

\bibitem[{\citenamefont{Gray et~al.}(2011)\citenamefont{Gray, Karel, Min\'ar,
  Bordel, Ebert, Braun, Ueda, Yamashita, Ouyang, Smith et~al.}}]{Gray2011}
\bibinfo{author}{\bibfnamefont{A.~X.} \bibnamefont{Gray}},
  \bibinfo{author}{\bibfnamefont{J.}~\bibnamefont{Karel}},
  \bibinfo{author}{\bibfnamefont{J.}~\bibnamefont{Min\'ar}},
  \bibinfo{author}{\bibfnamefont{C.}~\bibnamefont{Bordel}},
  \bibinfo{author}{\bibfnamefont{H.}~\bibnamefont{Ebert}},
  \bibinfo{author}{\bibfnamefont{J.}~\bibnamefont{Braun}},
  \bibinfo{author}{\bibfnamefont{S.}~\bibnamefont{Ueda}},
  \bibinfo{author}{\bibfnamefont{Y.}~\bibnamefont{Yamashita}},
  \bibinfo{author}{\bibfnamefont{L.}~\bibnamefont{Ouyang}},
  \bibinfo{author}{\bibfnamefont{D.~J.} \bibnamefont{Smith}},
  \bibnamefont{et~al.}, \bibinfo{journal}{Phys. Rev. B}
  \textbf{\bibinfo{volume}{83}}, \bibinfo{pages}{195112}
  (\bibinfo{year}{2011}).

\bibitem[{\citenamefont{Takahashi et~al.}(2008)\citenamefont{Takahashi,
  Igarashi, and Hamada}}]{Taka2008FeXPS}
\bibinfo{author}{\bibfnamefont{M.}~\bibnamefont{Takahashi}},
  \bibinfo{author}{\bibfnamefont{J.}~\bibnamefont{Igarashi}}, \bibnamefont{and}
  \bibinfo{author}{\bibfnamefont{N.}~\bibnamefont{Hamada}},
  \bibinfo{journal}{Phys. Rev. B} \textbf{\bibinfo{volume}{78}},
  \bibinfo{pages}{155108} (\bibinfo{year}{2008}).

\bibitem[{\citenamefont{Takahashi and Igarashi}(2010)}]{Taka2010FeCoNi3sXPS}
\bibinfo{author}{\bibfnamefont{M.}~\bibnamefont{Takahashi}} \bibnamefont{and}
  \bibinfo{author}{\bibfnamefont{J.-i.} \bibnamefont{Igarashi}},
  \bibinfo{journal}{Phys. Rev. B} \textbf{\bibinfo{volume}{81}},
  \bibinfo{pages}{035118} (\bibinfo{year}{2010}).

\bibitem[{\citenamefont{de~Groot and Kotani}(2008)}]{deGrootKotani}
\bibinfo{author}{\bibfnamefont{F.}~\bibnamefont{de~Groot}} \bibnamefont{and}
  \bibinfo{author}{\bibfnamefont{A.}~\bibnamefont{Kotani}},
  \emph{\bibinfo{title}{Core Level Spectroscopy of Solids}}
  (\bibinfo{publisher}{CRC Press, Boca Raton}, \bibinfo{year}{2008}).

\bibitem[{\citenamefont{Tanaka et~al.}(1992)\citenamefont{Tanaka, Jo, and
  Sawatzky}}]{Tanaka1992b}
\bibinfo{author}{\bibfnamefont{A.}~\bibnamefont{Tanaka}},
  \bibinfo{author}{\bibfnamefont{T.}~\bibnamefont{Jo}}, \bibnamefont{and}
  \bibinfo{author}{\bibfnamefont{G.~A.} \bibnamefont{Sawatzky}},
  \bibinfo{journal}{J. Phys. Soc. Jpn.} \textbf{\bibinfo{volume}{61}},
  \bibinfo{pages}{2636 } (\bibinfo{year}{1992}).

\bibitem[{\citenamefont{Mahan}(1980)}]{Mahan.PRB.21.1421}
\bibinfo{author}{\bibfnamefont{G.~D.} \bibnamefont{Mahan}},
  \bibinfo{journal}{Phys. Rev. B} \textbf{\bibinfo{volume}{21}},
  \bibinfo{pages}{1421} (\bibinfo{year}{1980}).

\bibitem[{\citenamefont{von Barth and Grossmann}(1979)}]{vonBarth1979}
\bibinfo{author}{\bibfnamefont{U.}~\bibnamefont{von Barth}} \bibnamefont{and}
  \bibinfo{author}{\bibfnamefont{G.}~\bibnamefont{Grossmann}},
  \bibinfo{journal}{Solid State Commun.} \textbf{\bibinfo{volume}{32}},
  \bibinfo{pages}{645 } (\bibinfo{year}{1979}).

\bibitem[{\citenamefont{Friedel}(1958)}]{Friedel1958}
\bibinfo{author}{\bibfnamefont{J.}~\bibnamefont{Friedel}},
  \bibinfo{journal}{Nuovo Cim.\ Suppl.} \textbf{\bibinfo{volume}{2}},
  \bibinfo{pages}{287} (\bibinfo{year}{1958}).

\bibitem[{\citenamefont{Hybertsen and Louie}(1986)}]{Hybertsen1986}
\bibinfo{author}{\bibfnamefont{M.~S.} \bibnamefont{Hybertsen}}
  \bibnamefont{and} \bibinfo{author}{\bibfnamefont{S.~G.} \bibnamefont{Louie}},
  \bibinfo{journal}{Phys. Rev. B} \textbf{\bibinfo{volume}{34}},
  \bibinfo{pages}{5390} (\bibinfo{year}{1986}).

\bibitem[{\citenamefont{Hamada et~al.}(1990)\citenamefont{Hamada, Hwang, and
  Freeman}}]{Hamada1990}
\bibinfo{author}{\bibfnamefont{N.}~\bibnamefont{Hamada}},
  \bibinfo{author}{\bibfnamefont{M.}~\bibnamefont{Hwang}}, \bibnamefont{and}
  \bibinfo{author}{\bibfnamefont{A.~J.} \bibnamefont{Freeman}},
  \bibinfo{journal}{Phys. Rev. B} \textbf{\bibinfo{volume}{41}},
  \bibinfo{pages}{3620} (\bibinfo{year}{1990}).

\bibitem[{\citenamefont{Kotani and Toyozawa}(1974)}]{Kotani1974}
\bibinfo{author}{\bibfnamefont{A.}~\bibnamefont{Kotani}} \bibnamefont{and}
  \bibinfo{author}{\bibfnamefont{Y.}~\bibnamefont{Toyozawa}},
  \bibinfo{journal}{J.\ Phys.\ Soc.\ Jpn.} \textbf{\bibinfo{volume}{37}},
  \bibinfo{pages}{912} (\bibinfo{year}{1974}).

\bibitem[{\citenamefont{Abrikosov et~al.}(2001)\citenamefont{Abrikosov,
  Olovsson, and Johansson}}]{Olovsson.PRL.87.176403}
\bibinfo{author}{\bibfnamefont{I.~A.} \bibnamefont{Abrikosov}},
  \bibinfo{author}{\bibfnamefont{W.}~\bibnamefont{Olovsson}}, \bibnamefont{and}
  \bibinfo{author}{\bibfnamefont{B.}~\bibnamefont{Johansson}},
  \bibinfo{journal}{Phys. Rev. Lett.} \textbf{\bibinfo{volume}{87}},
  \bibinfo{pages}{176403} (\bibinfo{year}{2001}).

\bibitem[{\citenamefont{Olovsson et~al.}(2005)\citenamefont{Olovsson,
  G\"oransson, Pourovskii, Johansson, and Abrikosov}}]{Olovsson.PRB.72.064203}
\bibinfo{author}{\bibfnamefont{W.}~\bibnamefont{Olovsson}},
  \bibinfo{author}{\bibfnamefont{C.}~\bibnamefont{G\"oransson}},
  \bibinfo{author}{\bibfnamefont{L.~V.} \bibnamefont{Pourovskii}},
  \bibinfo{author}{\bibfnamefont{B.}~\bibnamefont{Johansson}},
  \bibnamefont{and} \bibinfo{author}{\bibfnamefont{I.~A.}
  \bibnamefont{Abrikosov}}, \bibinfo{journal}{Phys. Rev. B}
  \textbf{\bibinfo{volume}{72}}, \bibinfo{pages}{064203}
  (\bibinfo{year}{2005}).

\bibitem[{\citenamefont{Akai}(1982)}]{Akai1982}
\bibinfo{author}{\bibfnamefont{H.}~\bibnamefont{Akai}}, \bibinfo{journal}{J.\
  Phys.\ Soc.\ Jpn.} \textbf{\bibinfo{volume}{51}}, \bibinfo{pages}{468}
  (\bibinfo{year}{1982}).

\bibitem[{\citenamefont{Akai}(1998)}]{Akai1998}
\bibinfo{author}{\bibfnamefont{H.}~\bibnamefont{Akai}},
  \bibinfo{journal}{Phys.\ Rev.\ Lett.} \textbf{\bibinfo{volume}{81}},
  \bibinfo{pages}{3002} (\bibinfo{year}{1998}).

\bibitem[{\citenamefont{Morruzi et~al.}(1978)\citenamefont{Morruzi, Janak, and
  Williams}}]{Morruzi1978}
\bibinfo{author}{\bibfnamefont{V.~L.} \bibnamefont{Morruzi}},
  \bibinfo{author}{\bibfnamefont{J.~F.} \bibnamefont{Janak}}, \bibnamefont{and}
  \bibinfo{author}{\bibfnamefont{A.~R.} \bibnamefont{Williams}},
  \emph{\bibinfo{title}{Calculated Electronic Properties of Metals}}
  (\bibinfo{publisher}{Pergamon, New York}, \bibinfo{year}{1978}).

\bibitem[{\citenamefont{Klebanoff et~al.}(1994)\citenamefont{Klebanoff,
  Van~Campen, and Pouliot}}]{Klebanoff1994}
\bibinfo{author}{\bibfnamefont{L.~E.} \bibnamefont{Klebanoff}},
  \bibinfo{author}{\bibfnamefont{D.~G.} \bibnamefont{Van~Campen}},
  \bibnamefont{and} \bibinfo{author}{\bibfnamefont{R.~J.}
  \bibnamefont{Pouliot}}, \bibinfo{journal}{Phys. Rev. B}
  \textbf{\bibinfo{volume}{49}}, \bibinfo{pages}{2047} (\bibinfo{year}{1994}).

\bibitem[{\citenamefont{See and Klebanoff}(1995)}]{See1995Ni}
\bibinfo{author}{\bibfnamefont{A.~K.} \bibnamefont{See}} \bibnamefont{and}
  \bibinfo{author}{\bibfnamefont{L.~E.} \bibnamefont{Klebanoff}},
  \bibinfo{journal}{Phys. Rev. B} \textbf{\bibinfo{volume}{51}},
  \bibinfo{pages}{11002} (\bibinfo{year}{1995}).

\bibitem[{\citenamefont{Braicovich and van~der
  Laan}(2008)}]{Braicovich.PRB.78.174421}
\bibinfo{author}{\bibfnamefont{L.}~\bibnamefont{Braicovich}} \bibnamefont{and}
  \bibinfo{author}{\bibfnamefont{G.}~\bibnamefont{van~der Laan}},
  \bibinfo{journal}{Phys. Rev. B} \textbf{\bibinfo{volume}{78}},
  \bibinfo{pages}{174421} (\bibinfo{year}{2008}).

\bibitem[{\citenamefont{Kruger and Natoli}(2004)}]{Kruger2004}
\bibinfo{author}{\bibfnamefont{P.}~\bibnamefont{Kruger}} \bibnamefont{and}
  \bibinfo{author}{\bibfnamefont{C.~R.} \bibnamefont{Natoli}},
  \bibinfo{journal}{Phys.\ Rev.\ B} \textbf{\bibinfo{volume}{70}},
  \bibinfo{pages}{245120} (\bibinfo{year}{2004}).

\end{thebibliography}

\end{document}